\documentstyle[preprint,aps]{revtex}

\newtheorem{theorem}{Theorem}
\newtheorem{acknowledgement}[theorem]{Acknowledgement}

\begin{document}
\title{Calculation of Expectation Values of S$_{z}$ and S$^{2}$ operators for
spin-1 and spin-3/2 particles}
\author{S. G\"{o}nen$^{\dagger }$ and A. Havare$^{\ddagger }$}
\address{$^{\dagger }$University of Dicle, Faculty of Education Science, Diyarbak\i\\
r, Turkey.\\
$^{\ddagger }$University of Mersin, Department of Physics, Mersin, Turkey}
\maketitle

\begin{abstract}
In the present article we calculate the expectation values of of S$_{z}$ and
S$^{2}$ operators for spin-1 and spin-3/2 particles by expanding a general
wave function which includes all spin values. The results are same as in the
stantard quantum mechanics.
\end{abstract}

\section{Introduction}

The electron is a fundemantal particle in the nature and its understanding
is based on the construction of electrodynamics. \ First studies concerning
electrons were made by Lorentz at 1905's. Dirac contributed these studies
extended Lorentz equation to contain radiative processes, and is now known
as the Lorentz-Dirac equation\cite{1,2}. In these studies, the electron was
considered in classical, point-like spinless particle. This classical
picture changed with the development of quantum mechanics giving new degrees
of freedom to the electron, the fundamental among which being spin\cite{3}.
In his equation, Pauli considered the electron as a non-relativistic
particle and quantum spin described with his exclusion principle\cite{4}. In
the two quantum mechanical descriptions proposed by Schr\"{o}dinger, one
assumes the electron to be a spinless point-like particle with
non-relativistic energies and described by Schr\"{o}dinger equation, while
the other allows relativistic energies and described by the Klein-Gordon
equation. However the best description of the electron is given by the Dirac
equation\cite{5}. All of these equations contains a spin concept which can
not be understood by classical mechanics and it is not obvious how the
correspondence principle would operate.

In recent years, there has been an increasing effort to understand both the
classical meaning of spin and the behaviour of a classical system according
to the rules of quantum mechanics\cite{6,7,8,9,10}. In these works, in
addition to usual the coordinate and momentum degrees of freedom, the
electron's internal degrees of freedom is also included. Though a classical
system both orbital momentum and spin angular momentum may have any value,
in quantum mechanics spin angular momentum may assume only integer multiples
of $\hbar /2$\cite{11}.Similar situations apply in vector-field equations of 
$\ \hbar $-spin$\,$\ and Rarita-Schwinger equations of $3\hbar /2$-spin\cite
{12}.

In this work we obtained a wave function by quantizing the classical model
and then expand this function in terms of the eigenvalues of spin operator.
Finally it is seen that this quantized wave function, which has spin
eigenvalues $0,1/2,1,3/2,2,...,n/2$ is a generalized wave function. It is
seen that from this quantized wave function, we can obtain the expectation
values of $S_{z}$ and $S^{2}$ operators for spin-1 and spin-3/2 particles..

\section{Equation of Motion for the Classical System}

The action of a classical spinning particle is given by Barut and Zhanghi 
\cite{7} as 
\begin{equation}
I=\int d\tau \left[ i\overline{z}\stackrel{\cdot }{z}+P_{\mu }\stackrel{%
\cdot }{x}^{\mu }-\pi _{\mu }\overline{z}\gamma ^{\mu }z\right] .  \label{1}
\end{equation}
Here, $x^{\mu }$ are the space-time coordinates, $z,i\overline{z}\in C^{4}$
are internal dynamics variables, and $\pi _{\mu }$ are general momentum. In
this case, the complete phase space has four sets of variables: $\left(
x^{\mu },P_{\mu },z,i\overline{z}\right) .$ We use a proper time formalism
as $z=z(\tau )$ and $x^{\mu }=x^{\mu }(\tau ),$ where $\tau $ being the
proper time of center of mass. Altough the mass term does not enter into the
Lagrangian it exist due to integral motion. The classical theory already
contains the notion of anti-particles in the form of positive and negative
internal frequencies for a given momentum P.

The equations of motion for the classical system are 
\begin{eqnarray}
\stackrel{\cdot }{z}\, &=&-i\pi _{\mu }\gamma ^{\mu }z  \label{2} \\
\stackrel{\cdot }{\overline{z}} &=&i\overline{z}\gamma ^{\mu }\pi _{\mu }
\label{3} \\
\stackrel{\cdot }{x}^{\mu } &=&v^{\mu }\overline{z}\gamma ^{\mu }z  \label{4}
\\
P^{\mu } &=&eA_{\alpha ,\mu }v^{\alpha }-\pi _{\alpha }\overline{z}\gamma
_{\mu }^{\alpha }z  \label{5}
\end{eqnarray}
and the Hamiltonian of the classical system, which is equal to the proper
mass, takes the form 
\begin{equation}
H=iz\overline{z}+P_{\mu }\stackrel{\cdot }{x}^{\mu }-L=\pi _{\mu }\overline{z%
}\gamma ^{\mu }z.  \label{6}
\end{equation}

The variation of the Hamiltonian is 
\begin{equation}
\delta H=\frac{\partial H}{\partial (i\overline{z})}\delta (i\overline{z})+%
\frac{\partial H}{\partial P_{\mu }}\delta P_{\mu }+\frac{\partial H}{%
\partial z}\delta z+\frac{\partial H}{\partial x^{\mu }}\delta x^{\mu }+%
\frac{\partial H}{\partial \tau }\delta \tau .  \label{7}
\end{equation}
We obtain Hamilton equations of motion from the Hamiltonian in Eq.(\ref{6})
as 
\begin{eqnarray}
\frac{\partial H}{\partial P_{\mu }} &=&\frac{dx^{\mu }}{d\tau },  \label{8}
\\
\frac{\partial H}{\partial (i\overline{z})} &=&\frac{dz}{d\tau },  \label{9}
\\
\frac{\partial H}{\partial z} &=&-\frac{d(i\overline{z})}{d\tau },
\label{10} \\
\frac{\partial H}{\partial x^{\mu }} &=&-\frac{dP^{\mu }}{d\tau }.
\label{11}
\end{eqnarray}
We can rewrite Eq.(\ref{7}) by using this equations of motion as 
\begin{equation}
\delta H=\frac{dz}{d\tau }\delta (i\overline{z})+\frac{dx^{\mu }}{d\tau }%
\delta P_{\mu }-\frac{d(i\overline{z})}{d\tau }\delta z-\frac{dP^{\mu }}{%
d\tau }\delta x^{\mu }+\frac{\partial H}{\partial \tau }\delta \tau .
\label{12}
\end{equation}
Using this expression we can write variation of action in Eq.(\ref{1}) as 
\begin{equation}
\delta I=(P_{\mu }dx^{\mu }+i\overline{z}dz-Hd\tau ).  \label{13}
\end{equation}

\section{Quantization}

The phase function, which is defined by quantum mechanics, depends on
classical action as 
\begin{equation}
S=\frac{1}{\hbar }I.  \label{14}
\end{equation}
\newline
The Hamilton-Jacobi, or phase function S, describes a family of possible
classical trajectories $(z(\tau ),x^{\mu }(\tau ))$ with momentum $P_{A}=(i%
\overline{z},P_{\mu }).$ In quantum theory, these possible classical
trajectories are the rays which belong to a wave with a definite momentum $%
P_{A}$ and proper mass $H=m$ described by the wave function as $(\hbar =1)$%
\begin{equation}
\Psi _{P}(z,x;\tau )=Ae^{iS(z,x;\tau )},  \label{15}
\end{equation}
where 
\begin{equation}
S(z,x^{\mu };\tau )=\left\{ 
\begin{array}{c}
\int (i\overline{z}dz+P_{\mu }dx^{\mu }-Hd\tau );\,\,P_{A}\text{ depends on }%
z,x^{\mu } \\ 
i\overline{z}z+P_{\mu }x^{\mu }-H\tau \text{ \ \ \ \ \ \ \ \ \ \ };\,\,i%
\overline{z},P_{\mu }\text{ are constants}
\end{array}
\right.  \label{16}
\end{equation}
By the substitution of $S$ (in the special case of constants $+i\overline{z}%
,P_{\mu })$ into Eq.(\ref{15}), it takes the form 
\begin{equation}
\Psi _{P_{A}}(z,x;\tau )=Ae^{i\left( +i\overline{z}z+P_{\mu }x^{\mu }-H\tau
\right) }.  \label{17}
\end{equation}
The wave function that represents the particle will be a superposition of
eigenfunctions. This general situation must be solution of Schr\"{o}dinger
wave equation. For a given wave function $\Psi (\overline{z},x;\tau )$
covariant Schr\"{o}dinger equation is 
\begin{equation}
i\frac{\partial \Psi }{\partial \tau }=H\Psi ,  \label{18}
\end{equation}
where $\overline{z}$ is the canonical conjugate of $z.$ It is possible to
obtain excited states of Zitterbewegung by expanding a general wave function 
$\Psi (\overline{z},x;\tau )$ power series of $\overline{z}$ \cite{13}: 
\[
\Psi (\overline{z},x;\tau )=\psi (x;\tau )+\frac{1}{1!}\overline{z}^{\alpha
}\psi _{\alpha }(x;\tau )+\frac{1}{2!}\overline{z}^{\alpha }\overline{z}%
^{\beta }\psi _{\alpha \beta }(x;\tau )+ 
\]
\begin{equation}
\frac{1}{3!}\overline{z}^{\alpha }\overline{z}^{\beta }\overline{z}^{\nu
}\psi _{\alpha \beta \nu }(x;\tau )+\cdots  \label{19}
\end{equation}
If we substitute this expansion into Eq.(\ref{18}) and separate the terms in
ordered form we get 
\begin{eqnarray}
i\frac{\partial }{\partial \tau }[\psi (x;\tau )] &=&\overline{z}\gamma
^{\mu }z\pi _{\mu }\psi _{\alpha }(x;\tau ),  \label{20} \\
i\frac{\partial }{\partial \tau }[\overline{z}^{\alpha }\psi _{\alpha
}(x;\tau )] &=&\overline{z}^{\eta }\gamma _{(\eta \sigma )}^{\mu }z^{\sigma
}\pi _{\mu }\overline{z}^{\alpha }\psi _{\alpha }(x;\tau ),  \label{21} \\
i\frac{\partial }{\partial \tau }[\overline{z}^{\alpha }\overline{z}^{\beta
}\psi _{\alpha \beta }(x;\tau )] &=&\overline{z}^{\eta }\gamma _{(\eta
\sigma )}^{\mu }z^{\sigma }\pi _{\mu }\overline{z}^{\alpha }\overline{z}%
^{\beta }\psi _{\alpha \beta }(x;\tau ),  \label{22} \\
i\frac{\partial }{\partial \tau }[\overline{z}^{\alpha }\overline{z}^{\beta }%
\overline{z}^{\nu }\psi _{\alpha \beta \nu }(x;\tau )] &=&\overline{z}^{\eta
}\gamma _{(\eta \sigma )}^{\mu }z^{\sigma }\pi _{\mu }\overline{z}^{\alpha }%
\overline{z}^{\beta }\overline{z}^{\nu }\psi _{\alpha \beta \nu }(x;\tau ),
\label{23} \\
&&\vdots  \nonumber \\
i\frac{\partial }{\partial \tau }[\overline{z}^{\alpha _{1}}\overline{z}%
^{\alpha _{2}}\ldots \overline{z}^{\alpha _{n}}\psi _{\alpha _{1},\alpha
_{2},\ldots \alpha _{n}}(x;\tau )] &=&\overline{z}^{\eta }\gamma _{(\eta
\sigma )}^{\mu }z^{\sigma }\pi _{\mu }\overline{z}^{\alpha _{1}}\overline{z}%
^{\alpha _{2}}\ldots \overline{z}^{\alpha _{n}}\psi _{\alpha _{1},\alpha
_{2},\ldots \alpha _{n}}(x;\tau ).  \label{24}
\end{eqnarray}
Starting from Eq.(\ref{20}) we can obtain explicit form of these equations.
It is known that from Euler-Lagrange equations of classical system 
\begin{equation}
\stackrel{.}{x}^{\mu }=v^{\mu }=\overline{z}\gamma ^{\mu }z.  \label{25}
\end{equation}
Using this and introducing wave function $\psi (x;\tau )$ as 
\begin{equation}
\psi (x;\tau )=U(x)e^{-im\tau }  \label{26}
\end{equation}
Eq.(\ref{20}) takes the form 
\begin{equation}
\lbrack v^{\mu }\pi _{\mu }-m]U(x)=0.  \label{27}
\end{equation}
In the spinless case it is seen that 
\begin{equation}
v^{\mu }v_{\mu }=1  \label{28}
\end{equation}
and 
\begin{equation}
\pi ^{\mu }\pi _{\mu }=m^{2}.  \label{29}
\end{equation}
Comparing these two conditions we obtain 
\begin{equation}
v^{\mu }=\frac{1}{m}\pi ^{\mu }.  \label{30}
\end{equation}
Substituting Eq.(\ref{30}) in Eq.(\ref{27}) we get the following equation,
and is known as Klein-Gordon equation: 
\begin{equation}
\lbrack \pi ^{\mu }\pi _{\mu }-m^{2}]U(x)=0.  \label{31}
\end{equation}
In Eq.(\ref{21}), if choose the dynamics variable $\overline{z}^{\alpha }$
as time independent and use Eq.(\ref{26}) we obtain 
\begin{equation}
\lbrack \gamma ^{\mu }\pi _{\mu }-m^{2}]U(x)=0.  \label{32}
\end{equation}
This is known as the Dirac equation describing spin-$1/2$ particles.

Similarly for Eq.(\ref{22}) we can simplify calculations using 
\begin{equation}
\overline{z}^{\alpha }\overline{z}^{\beta }\psi _{\alpha \beta }(x;\tau )=%
\frac{1}{2}\overline{z}^{\alpha }\overline{z}^{\beta }(\psi _{\alpha \beta
}+\psi _{\beta \alpha }),  \label{33}
\end{equation}
where $\overline{z}^{\alpha }\overline{z}^{\beta }=\overline{z}^{\beta }%
\overline{z}^{\alpha }.$ Then Eq.(\ref{22}) takes the form 
\begin{equation}
i\frac{\partial }{\partial \tau }[\psi _{\alpha \beta }+\psi _{\beta \alpha
}]=[\gamma _{(\alpha \beta )}^{\mu }\otimes \text{I}+\text{I}\otimes \gamma
_{(\alpha \beta )}^{\mu }]\pi ^{\mu }(\psi _{\alpha \beta }+\psi _{\beta
\alpha }).  \label{34}
\end{equation}
Here, if we introduce 
\begin{equation}
\Psi _{\alpha \beta }=(\psi _{\alpha \beta }+\psi _{\beta \alpha })
\label{35}
\end{equation}
and 
\begin{equation}
\beta ^{\mu }=[\gamma _{(\alpha \beta )}^{\mu }\otimes \text{I}+\text{I}%
\otimes \gamma _{(\alpha \beta )}^{\mu }],  \label{36}
\end{equation}
and use Eq.(\ref{26}), Eq.(\ref{34}) gives 
\begin{equation}
\lbrack \beta ^{\mu }\pi _{\mu }-m]U_{\alpha \beta }(x)=0.  \label{37}
\end{equation}
This is the Kemmer equation and describes spin-1 particles. The $\beta ^{\mu
}$ are $16\times 16$ Hermitian matrices and are known as Kemmer matrices.

If we use in Eq.(\ref{23}) the following expression 
\begin{equation}
\overline{z}^{\alpha }\overline{z}^{\beta }\overline{z}^{\nu }\psi _{\alpha
\beta \nu }(x;\tau )=\frac{1}{6}\overline{z}^{\alpha }\overline{z}^{\beta }%
\overline{z}^{\nu }(\psi _{\alpha \beta \nu }+\psi _{\alpha \nu \beta }+\psi
_{\beta \alpha \nu }+\psi _{\beta \nu \alpha }+\psi _{\nu \alpha \beta
}+\psi _{\nu \beta \alpha })  \label{38}
\end{equation}
we get 
\begin{equation}
i\frac{\partial }{\partial \tau }\Psi _{\alpha \beta \nu }=[\gamma ^{\mu
}\otimes \text{I}_{\alpha \nu }\otimes \text{I}_{\nu \alpha }+\text{I}%
_{\beta \nu }\otimes \gamma ^{\mu }\otimes \text{I}_{\nu \beta }+\text{I}%
_{\alpha \nu }\otimes \text{I}_{\nu \alpha }\otimes \gamma ^{\mu }]\pi _{\mu
}\Psi _{\alpha \beta \nu }.  \label{39}
\end{equation}
If we introduce 
\begin{eqnarray}
\alpha ^{\mu } &=&[\gamma ^{\mu }\otimes \text{I}_{\alpha \nu }\otimes \text{%
I}_{\nu \alpha }+\text{I}_{\beta \nu }\otimes \gamma ^{\mu }\otimes \text{I}%
_{\nu \beta }+\text{I}_{\alpha \nu }\otimes \text{I}_{\nu \alpha }\otimes
\gamma ^{\mu }],  \label{40} \\
\Psi _{\alpha \beta \nu } &=&\frac{1}{6}(\psi _{\alpha \beta \nu }+\psi
_{\alpha \nu \beta }+\psi _{\beta \alpha \nu }+\psi _{\beta \nu \alpha
}+\psi _{\nu \alpha \beta }+\psi _{\nu \beta \alpha }),  \label{41}
\end{eqnarray}
and use Eq.(\ref{26}) we obtain the following equation which is known as
Rarita-Schwinger equation describing spin-3/2 particles: 
\begin{equation}
\lbrack \alpha ^{\mu }\pi _{\mu }-m]U_{\alpha \beta \nu }(x)=0.  \label{42}
\end{equation}
Now, general term Eq.(\ref{24}) gives 
\begin{equation}
\lbrack \alpha ^{\lambda }\pi _{\lambda }-m]U_{\alpha _{1},\alpha
_{2},\ldots \alpha _{n}}(x)=0,  \label{43}
\end{equation}
where 
\begin{equation}
\alpha ^{\lambda }=[\gamma ^{\lambda }\otimes \text{I}\otimes \ldots \otimes 
\text{I}+\text{I}\otimes \gamma ^{\lambda }\otimes \ldots \otimes \text{I}+%
\text{I}\otimes \text{I}\otimes \ldots \otimes \gamma ^{\lambda }].
\label{44}
\end{equation}
These $n\times n$ matrices satisfies 
\begin{equation}
\mathrel{\mathop{\sum }\limits_{(P)}}%
\alpha _{\mu _{1}}\ldots \alpha _{\mu _{L}}(\alpha _{\mu _{L}+1}\alpha _{\mu
_{L}+2}-\delta _{\mu _{L}+1,\mu _{L}+2})=0,  \label{45}
\end{equation}
where the sum $\sum\limits_{(P)}$ is performed over all possible
permutations of $\mu _{1},\mu _{2},\ldots ,$ $\mu _{L+1},\mu _{L+2\text{ }}$%
indices. Equation (\ref{43}) describes the dynamics of spin-$n/2$ particles.

\section{\protect\bigskip Calculation The Expectation Values of S$_{Z}$ and S%
$^{2}$ for Spin-1 Particle}

The series expansion terms of the wave function in terms of the internal
variables describe the spin-$0$, $\frac{1}{2}$, $1$, $\frac{3}{2}$, ..., $%
\frac{n}{2}$ states respectively. Here, we only discuss the spin-$1$ and
spin-$\frac{3}{2}$ particle states.

From the definition of the spin tensor 
\begin{equation}
S_{\mu \upsilon }=\frac{i}{4}\overline{z}\left[ \gamma _{\mu },\gamma
_{\upsilon }\right] z
\end{equation}
we can write 
\begin{equation}
S_{z}=\frac{1}{2}\left( \overline{z}_{+}z_{+}-\overline{z}_{-}z_{-}\right) 
\end{equation}
where $\mu ,\upsilon =0,1,2,3$ . $\gamma _{\mu },\gamma _{\upsilon }$ are
usual Dirac matrices and 
\begin{eqnarray}
z_{+} &=&\frac{\partial }{\partial \overline{z}_{+}} \\
z_{-} &=&\frac{\partial }{\partial \overline{z}_{-}}\text{ \ .}
\end{eqnarray}
With the aid of the operator definitions we can write 
\begin{equation}
S_{z}=\frac{1}{2}\left( \overline{z}_{+}\frac{\partial }{\partial \overline{z%
}_{+}}-\overline{z}_{-}\frac{\partial }{\partial \overline{z}_{-}}\right) 
\end{equation}
Considering the spin orientations if we define the wave function of the
spin-1 particle as 
\begin{equation}
\overline{z}_{\alpha }\overline{z}_{\beta }\Psi ^{\alpha \beta }\left(
x;\tau \right) =\left\{ \overline{z}_{+}\overline{z}_{+}\Psi ^{++}\left(
x;\tau \right) +\overline{z}_{+}\overline{z}_{-}(\Psi ^{+-}\left( x;\tau
\right) +\Psi ^{-+}\left( x;\tau \right) )+\overline{z}_{-}\overline{z}%
_{-}\Psi ^{--}\left( x;\tau \right) \right\} 
\end{equation}
we find the these expectation values; 
\[
\left\langle S_{z}\right\rangle _{+}=\langle \overline{z}_{+}\overline{z}%
_{+}\Psi ^{++}\left( x;\tau \right) \left| \frac{1}{2}\left( \overline{z}_{+}%
\frac{\partial }{\partial \overline{z}_{+}}-\overline{z}_{-}\frac{\partial }{%
\partial \overline{z}_{-}}\right) \right| 
\]
\begin{equation}
\overline{z}_{+}\overline{z}_{+}\Psi ^{++}\left( x;\tau \right) \rangle =1
\end{equation}

\[
\left\langle S_{z}\right\rangle _{0}=\langle \overline{z}_{+}\overline{z}%
_{-}(\Psi ^{+-}\left( x;\tau \right) +\Psi ^{-+}\left( x;\tau \right) ) 
\]
\[
\left| \frac{1}{2}\left( \overline{z}_{+}\frac{\partial }{\partial \overline{%
z}_{+}}-\overline{z}_{-}\frac{\partial }{\partial \overline{z}_{-}}\right)
\right| 
\]
\begin{equation}
\overline{z}_{+}\overline{z}_{-}(\Psi ^{+-}\left( x;\tau \right) +\Psi
^{-+}\left( x;\tau \right) )\rangle =0
\end{equation}

\[
\left\langle S_{z}\right\rangle _{-}=\langle \overline{z}_{-}\overline{z}%
_{-}\Psi ^{--}\left( x;\tau \right) \left| \frac{1}{2}\left( \overline{z}_{+}%
\frac{\partial }{\partial \overline{z}_{+}}-\overline{z}_{-}\frac{\partial }{%
\partial \overline{z}_{-}}\right) \right| 
\]
\begin{equation}
\overline{z}_{-}\overline{z}_{-}\Psi ^{--}\left( x;\tau \right) \rangle =-1
\end{equation}
where +, 0, - subscripts describe the +1, 0 and -1 eigenvalues of the $S_{z%
\text{ }}$operator.

Then we can calculate the expectation value of the $S^{2}$ by using the wave
function of the spin-1 particle. As in the standart quantum mechanics we
first write the $S^{2}$ in terms of the annihilation and creation operators
as 
\begin{equation}
S^{2}=S_{+}S_{-}+S_{z}^{2}-S_{z}
\end{equation}
where 
\begin{eqnarray}
S_{-} &=&S_{x}-iS_{y} \\
S_{+} &=&S_{x}+iS_{y}\text{ \ .}
\end{eqnarray}
If we write every terms of $S^{2}$ in terms of internal variables, 
\begin{eqnarray}
S_{+} &=&\overline{z}_{+}z_{-} \\
S_{-} &=&\overline{z}_{-}z_{+} \\
S_{+}S_{-} &=&\overline{z}_{+}z_{+}+\overline{z}_{+}\overline{z}%
_{-}z_{-}z_{+} \\
S_{z}^{2} &=&\frac{1}{4}\left( \overline{z}_{+}z_{+}+\overline{z}_{+}%
\overline{z}_{+}z_{+}z_{+}+\overline{z}_{-}z_{-}+\overline{z}_{-}\overline{z}%
_{-}z_{-}z_{-}\right)
\end{eqnarray}
and use the Eq.(50) definition of $S_{z}$ we obtain the expectation value of 
$S^{2}$ as 
\[
\left\langle S^{2}\right\rangle =\langle \{\overline{z}_{+}\overline{z}%
_{+}\Psi ^{++}\left( x;\tau \right) +\overline{z}_{+}\overline{z}_{-}(\Psi
^{+-}\left( x;\tau \right) +\Psi ^{-+}\left( x;\tau \right) ) 
\]
\[
+\overline{z}_{-}\overline{z}_{-}\Psi ^{--}\left( x;\tau \right) \}\mid 
\overline{z}_{+}z_{+}+\overline{z}_{+}\overline{z}_{-}z_{-}z_{+}-\frac{1}{2}%
\left( \overline{z}_{+}z_{+}-\overline{z}_{-}z_{-}\right) +\frac{1}{4}(%
\overline{z}_{+}z_{+}+ 
\]
\[
\overline{z}_{+}\overline{z}_{+}z_{+}z_{+}+\overline{z}_{-}z_{-}+\overline{z}%
_{-}\overline{z}_{-}z_{-}z_{-})\mid \{\overline{z}_{+}\overline{z}_{+}\Psi
^{++}\left( x;\tau \right) + 
\]
\begin{equation}
\overline{z}_{+}\overline{z}_{-}(\Psi ^{+-}\left( x;\tau \right) +\Psi
^{-+}\left( x;\tau \right) )+\overline{z}_{-}\overline{z}_{-}\Psi
^{--}\left( x;\tau \right) \}\rangle =2
\end{equation}
This is the same value we obtain from the eigenvalue equation 
\begin{equation}
S^{2}\chi =s\left( s+1\right) \chi
\end{equation}
so, the $\overline{z}^{\alpha }\overline{z}^{\beta }\Psi _{\alpha \beta
}\left( x;\tau \right) $ wave function can be interpreted as an eigenstate
of eigenvalue 2 of $S^{2}$ operator.

\section{Calculation The Expectation Values of S$_{Z}$ and S$^{2}$ for Spin-$%
\frac{3}{2}$ Particle}

The third term of the series expansion of the general wave function $\Psi
\left( \overline{z},x;\tau \right) $ represants the spin-$\frac{3}{2}$
particle. This term is 
\begin{equation}
\overline{z}_{\alpha }\overline{z}_{\beta }\overline{z}_{\upsilon }\Psi
^{\alpha \beta \upsilon }\left( x;\tau \right) \text{.}
\end{equation}
$\overline{z}_{\alpha }\overline{z}_{\beta }\overline{z}_{\upsilon }$
corresponds to internal dynamics of the particle ultemitaly to spin motion.
Each $\overline{z}$ is related to spin with $\frac{1}{2}$ value. If we
choose the spin up orientations (+) and spin down orientations (-) then; 
\[
\overline{z}_{\alpha }\overline{z}_{\beta }\overline{z}_{\upsilon }\Psi
^{\alpha \beta \upsilon }\left( x;\tau \right) =\{\overline{z}_{+}\overline{z%
}_{+}\overline{z}_{+}\Psi ^{+++}\left( x;\tau \right) + 
\]
\[
\overline{z}_{+}\overline{z}_{+}\overline{z}_{-}(\Psi ^{++-}\left( x;\tau
\right) +\Psi ^{+-+}\left( x;\tau \right) +\Psi ^{-++}\left( x;\tau \right)
)+ 
\]
\[
\overline{z}_{+}\overline{z}_{-}\overline{z}_{-}(\Psi ^{+--}\left( x;\tau
\right) +\Psi ^{-+-}\left( x;\tau \right) +\Psi ^{--+}\left( x;\tau \right)
)+ 
\]
\begin{equation}
\overline{z}_{-}\overline{z}_{-}\overline{z}_{-}\Psi ^{---}\left( x;\tau
\right) \}
\end{equation}
Expectation values of $S_{z}$ due to spin orientations are 
\[
\left\langle S_{z}\right\rangle _{+++}=\langle \overline{z}_{+}\overline{z}%
_{+}\overline{z}_{+}\Psi ^{+++}\left( x;\tau \right) \left| \frac{1}{2}%
\left( \overline{z}_{+}\frac{\partial }{\partial \overline{z}_{+}}-\overline{%
z}_{-}\frac{\partial }{\partial \overline{z}_{-}}\right) \right| 
\]
\begin{equation}
\overline{z}_{+}\overline{z}_{+}\overline{z}_{+}\Psi ^{+++}\left( x;\tau
\right) \rangle =\frac{3}{2}
\end{equation}
\[
\left\langle S_{z}\right\rangle _{++-}=\langle \overline{z}_{+}\overline{z}%
_{+}\overline{z}_{-}(\Psi ^{++-}\left( x;\tau \right) +\Psi ^{+-+}\left(
x;\tau \right) +\Psi ^{-++}\left( x;\tau \right) ) 
\]
\[
\left| \frac{1}{2}\left( \overline{z}_{+}\frac{\partial }{\partial \overline{%
z}_{+}}-\overline{z}_{-}\frac{\partial }{\partial \overline{z}_{-}}\right)
\right| 
\]
\begin{equation}
\overline{z}_{+}\overline{z}_{+}\overline{z}_{-}(\Psi ^{++-}\left( x;\tau
\right) +\Psi ^{+-+}\left( x;\tau \right) +\Psi ^{-++}\left( x;\tau \right)
)\rangle =\frac{1}{2}
\end{equation}
\[
\left\langle S_{z}\right\rangle _{+--}=\langle \overline{z}_{+}\overline{z}%
_{-}\overline{z}_{-}(\Psi ^{+--}\left( x;\tau \right) +\Psi ^{-+-}\left(
x;\tau \right) +\Psi ^{--+}\left( x;\tau \right) ) 
\]
\[
\left| \frac{1}{2}\left( \overline{z}_{+}\frac{\partial }{\partial \overline{%
z}_{+}}-\overline{z}_{-}\frac{\partial }{\partial \overline{z}_{-}}\right)
\right| 
\]
\begin{equation}
\overline{z}_{+}\overline{z}_{-}\overline{z}_{-}(\Psi ^{+--}\left( x;\tau
\right) +\Psi ^{-+-}\left( x;\tau \right) +\Psi ^{--+}\left( x;\tau \right)
)\rangle =-\frac{1}{2}
\end{equation}
\[
\left\langle S_{z}\right\rangle _{---}=\langle \overline{z}_{-}\overline{z}%
_{-}\overline{z}_{-}\Psi ^{---}\left( x;\tau \right) \left| \frac{1}{2}%
\left( \overline{z}_{+}\frac{\partial }{\partial \overline{z}_{+}}-\overline{%
z}_{-}\frac{\partial }{\partial \overline{z}_{-}}\right) \right| 
\]
\begin{equation}
\overline{z}_{-}\overline{z}_{-}\overline{z}_{-}\Psi ^{---}\left( x;\tau
\right) \rangle =-\frac{3}{2}
\end{equation}
where $+++$, $++-$, $+--$, $---$ subscripts describe the eigenstates of +$%
\frac{3}{2}$, +$\frac{1}{2}$, -$\frac{1}{2}$ and -$\frac{3}{2}$ eigenvalues
of the $S_{z\text{ }}$operator. \ At the same time these results are the
projection of the spin vector on the z-axis due to $m_{s}=-s,-s+1,...,s-1,+s$
.

Again as in the spin-1 case we can also calculate the expectation value of
the $S^{2}$ by using the wave function of the spin-3/2 particle: 
\[
\left\langle S^{2}\right\rangle =\left\langle \overline{z}_{\alpha }%
\overline{z}_{\beta }\overline{z}_{\upsilon }\Psi ^{\alpha \beta \upsilon
}\left( x;\tau \right) \left| S^{2}\right| \overline{z}_{\alpha }\overline{z}%
_{\beta }\overline{z}_{\upsilon }\Psi ^{\alpha \beta \upsilon }\left( x;\tau
\right) \right\rangle 
\]
\[
=\langle \{\overline{z}_{+}\overline{z}_{+}\overline{z}_{+}\Psi ^{+++}\left(
x;\tau \right) +
\]
\[
\overline{z}_{+}\overline{z}_{+}\overline{z}_{-}(\Psi ^{++-}\left( x;\tau
\right) +\Psi ^{+-+}\left( x;\tau \right) +\Psi ^{-++}\left( x;\tau \right)
)+
\]
\[
\overline{z}_{+}\overline{z}_{-}\overline{z}_{-}(\Psi ^{+--}\left( x;\tau
\right) +\Psi ^{-+-}\left( x;\tau \right) +\Psi ^{--+}\left( x;\tau \right)
)+\overline{z}_{-}\overline{z}_{-}\overline{z}_{-}\Psi ^{---}\left( x;\tau
\right) \}
\]
\[
\mid \overline{z}_{+}z_{+}+\overline{z}_{+}\overline{z}_{-}z_{-}z_{+}-\frac{1%
}{2}\left( \overline{z}_{+}z_{+}-\overline{z}_{-}z_{-}\right) +\frac{1}{4}(%
\overline{z}_{+}z_{+}+
\]
\[
\overline{z}_{+}\overline{z}_{+}z_{+}z_{+}+\overline{z}_{-}z_{-}+\overline{z}%
_{-}\overline{z}_{-}z_{-}z_{-})\mid \{\overline{z}_{+}\overline{z}_{+}%
\overline{z}_{+}\Psi ^{+++}\left( x;\tau \right) +
\]
\[
\overline{z}_{+}\overline{z}_{+}\overline{z}_{-}(\Psi ^{++-}\left( x;\tau
\right) +\Psi ^{+-+}\left( x;\tau \right) +\Psi ^{-++}\left( x;\tau \right)
)+
\]
\[
\overline{z}_{+}\overline{z}_{-}\overline{z}_{-}(\Psi ^{+--}\left( x;\tau
\right) +\Psi ^{-+-}\left( x;\tau \right) +\Psi ^{--+}\left( x;\tau \right)
)+\overline{z}_{-}\overline{z}_{-}\overline{z}_{-}\Psi ^{---}\left( x;\tau
\right) \}\rangle 
\]
\begin{equation}
=\frac{15}{4}\text{ .}
\end{equation}
This is the same value we obtain from the eigenvalue equation 
\begin{equation}
S^{2}\chi =s\left( s+1\right) \chi 
\end{equation}
so, the $\overline{z}_{\alpha }\overline{z}_{\beta }\overline{z}_{\upsilon
}\Psi ^{\alpha \beta \upsilon }\left( x;\tau \right) $ wave function can be
interpreted as an eigenstate of eigenvalue $\frac{15}{4}$ of $S^{2}$
operator.

\section{Conclusion}

In this work, to understand the Schr\"{o}dinger equation, after quantizing
the classical system with spin, a generalized wave function was expanded
into a power series in terms of internal coordinates of the system. Then it
was shown that this expansion includes zero and positive integer values 
h\hskip-.2em\llap{\protect\rule[1.1ex]{.325em}{.1ex}}\hskip.2em%
/2, hence containing all fermion states (odd integer of 
h\hskip-.2em\llap{\protect\rule[1.1ex]{.325em}{.1ex}}\hskip.2em%
/2) and all boson states (even integer of 
h\hskip-.2em\llap{\protect\rule[1.1ex]{.325em}{.1ex}}\hskip.2em%
/2). Also by using the expanded wave function terms we calculated the
expectation values of S$_{z}$ and S$^{2}$ operators for spin-1 and spin-3/2
particles and saw that the results are in agrement with the results of
standart quantum mechaincs.

\begin{acknowledgement}
The authers thank to Prof Nuri Unal for helpfull discussions.
\end{acknowledgement}


\begin{references}
\bibitem{1}  H.A. Lorentz, ``The Theory of Electrons'', Dover, N. Y. (1952).

\bibitem{2}  P.A.M. Dirac, Proc. Roy. Soc. (London) {\bf A}167, 148(1958).

\bibitem{3}  H.A. Kramers, ``Quantum Mechanics'', Dover, N.Y.(1964).

\bibitem{4}  W. Pauli, Z. Phys., {\bf 31}, 765(1925).

\bibitem{5}  T-Y Wu and W-Y Pauchy Hwang, ``Relativistic Quantum Mechanics
and Quantum Fields'', World Scientific(1991).

\bibitem{6}  A.T. Ogelski and J. Sobcyzk, J. Math. Phys., {\bf 22}(1981).

\bibitem{7}  A.O. Barut and N. Zanghi, Phys. Rev. Lett., {\bf 52},
2009(1984).

\bibitem{8}  A.O. Barut, ``What is Electron? Relativistic Electron Theory
and Radiative Process'', Quantum Optics, Experimental Gravitation and
Measurement Theory (ed. P.Mystre and M.O. Scully), Plenum Press, N. Y.,
pp.155, (1983).

\bibitem{9}  A.O. Barut and I.H. Duru, Phys.Rev. Lett., {\bf 53},2355(1984).

\bibitem{10}  A.O. Barut and N. \"{U}nal, Phys. Rev.Lett. {\bf A40},
5404(1989).

\bibitem{11}  A.O. Barut, Phys. Rev. Lett. {\bf B},436(1990).

\bibitem{12}  Y. Takahashi, ``An Introduction to Field Quantization'',
Pergamon Press, London(1969).

\bibitem{13}  S. G\"{o}nen and A. Havare, Turk J. Phys., 13-17, 26 (2002).
\end{references}
\end{document}